\documentclass[global,referee]{svjour}

\usepackage{graphicx}
\usepackage{textcomp}
\usepackage{gensymb}
\usepackage[version=3,textfontname=sffamily]{mhchem} 
\usepackage[tight,nice]{units}
\usepackage{cite}
\usepackage{multirow}
\usepackage[english]{babel}

\journalname{Applied Physics B: Laser and Optics}

\let\olditemize=\itemize
\def\itemize{
\olditemize \setlength{\itemsep}{-0.5ex} }
\let\oldenumerate=\enumerate
\def\enumerate{
\oldenumerate \setlength{\itemsep}{-0.5ex} }

\newcommand{\muT}{\mbox{\micro T}}

\newcommand{\lampHg}{\ensuremath{{}^{204}\text{Hg}}}
\newcommand{\magHg}{\ensuremath{{}^{199}\text{Hg}}}

\graphicspath{{Image/}}

\begin{document}

\title{Experimental study of $^{199}$Hg spin anti-relaxation coatings}

\author{Z.~Chowdhuri\inst{1} \and
M.~Fertl\inst{1} \and
M.~Horras\inst{1} \and
K.~Kirch\inst{1,2} \and
J.~Krempel\inst{2} \and
B.~Lauss\inst{1} \and
A.~Mtchedlishvili\inst{1} \and
D.~Rebreyend\inst{3} \and
S.~Roccia\inst{4} \and
P.~Schmidt-Wellenburg\inst{1}\and
and G.~Zsigmond\inst{1}}

\institute{Paul Scherrer Institut, 5232 Villigen PSI, Switzerland  \and
ETH Z\"{u}rich, Switzerland\and
LPSC, Universit\'{e} Joseph Fourier Grenoble 1, CNRS/IN2P3, Grenoble INP, Grenoble, France\and
CSNSM, Universit\'{e} Paris Sud, CNRS/IN2P3, Orsay, France}

\mail{philipp.schmidt-wellenburg@psi.ch}

\date{Received: date / Revised version: date}

\maketitle

\begin{abstract}We report on a comparison of spin relaxation rates in a $^{199}$Hg magnetometer using different wall coatings.
A compact mercury magnetometer was built for this purpose. Glass cells coated with fluorinated materials show longer spin coherence times than if coated with their hydrogenated homologues. The longest spin relaxation time of the mercury vapor was measured with a fluorinated paraffin wall coating.
\end{abstract}


\section*{Introduction}
Magnetometers using optically pumped spin populations of atoms to measure magnetic fields
have achieved sensitivities comparable to most superconducting quantum interference devices (SQUID)-based magnetometers during the last decades\,\cite{Budker2007}.
Mercury magnetometers~(HgM), first used for searches of a \magHg{} electric dipole moment (EDM)\,\cite{Lamoreaux1987,Jacobs1993} and later as a cohabiting magnetometer in a neutron EDM~(nEDM) search\,\cite{Green1998}, are based on the optical detection of the Larmor precession with a frequency $\omega_L$ of spin-1/2 \magHg{} atoms in a magnetic field~$B$:

\begin{equation}
	\omega_L= \gamma_\text{Hg}\!\cdot\!B,
\label{eq:1}
\end{equation}

\noindent where $\gamma_\text{Hg}$ is the gyromagnetic factor of \magHg. An intrinsic advantage of measuring a free induction decay is that the precession frequency is
directly related to the absolute magnitude of the magnetic field. The sensitivity $\delta B$ of an optical magnetometer is proportional to $\left(\tau T N\right)^{-1/2}$, where $\tau$ is the spin coherence time also referred to as the depolarization time, $T$ the observation time, and $N$ is the number of atoms. 
However, in the case of a \magHg{} magnetometer we expect photon shot noise to dominate
the sensitivity $\delta B_\text{p}$ given by e.g.\ equation~(25) in Ref.\,\cite{Swallows2013} for an exponentially decaying sinusoidal signal. 
%
%
%
This means $\delta B_\text{p} \propto a_\text{n}/a_\text{s}$ where $a_\text{s}$ is the initial signal amplitude of the sine and $a_\text{n}$ is the RMS noise amplitude.
In the case of homogeneous fields the decay time constant $\tau$ is dominated by depolarization of \magHg{} atoms during collisions with the walls confining the atomic vapor.
Our study was motivated by magnetic field monitoring requirements needed for an improved search of a nEDM\,\cite{Baker2011} at the Paul Scherrer Institut~(PSI).
As the statistical sensitivity increases, the experiment also needs an improved control and monitoring of magnetic field stability to assess field drifts on a comparable sensitivity level.

In the PSI nEDM apparatus a batch of \magHg{} vapor is polarized in a dedicated cell, while the spin precession of neutrons and a previously prepared batch of mercury is measured in the adjacent main precession chamber.
At the start of each new measurement cycle the polarized
\magHg{} vapor is released into the precession chamber where a $\pi/2$-flip starts the Larmor precession.
Decreasing the depolarization rate due to wall collisions in the separate polarizing cell is a promising path to higher sensitivity as this will increase the equilibrium polarization before release, and hence the signal amplitude $a_\text{s}$.

In the past low viscosity Fomblin\footnote{Fomblin$^{\text{\textregistered}}$ is a registered trademark of Solvay Solexis.} oil of type ``Y'' was used to coat the glass polarization cell.
Over the course of weeks, the high surface tension of this type of Fomblin oil causes the film to pull away and expose sections of the bare glass.
Re-coating of the cell requires venting of the apparatus and the disassembly of the mercury magnetometer, a lengthy and tedious process. 
In order to find a better material which would minimize these regular interventions we decided to study alternative anti-relaxation coatings.
A dedicated mercury magnetometer was built at PSI\,\cite{Horras2012} to study depolarization times $\tau$ of several different wall coatings.

\begin{figure}
\centering
	\includegraphics[width=0.9\linewidth]{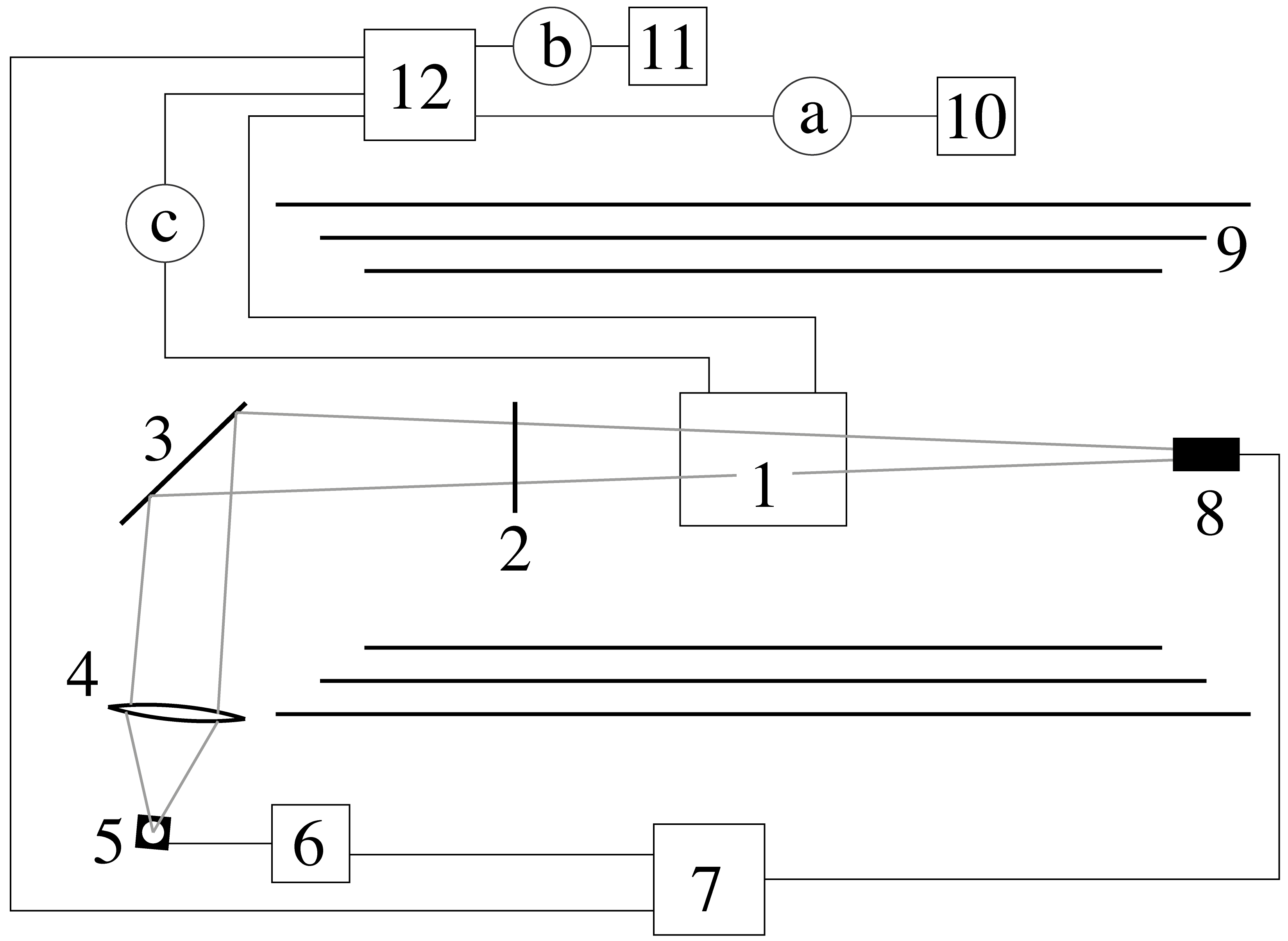}
	
	\caption{Schematic of the setup used to investigate several different materials as anti spin-relaxation coatings. 1)~Hg cell, 2)~$1/4$-wave plate, 3)~Brewster plate for linear polarization, 4)~lens, 5)~discharge \lampHg-bulb and microwave cavity, 6)~microwave generator, 7)~DAQ and control system, 8)~photo multiplier tube, 9)~magnetic shield, 10)~\magHg-source, 11)~vacuum pump, 12)~vacuum and Hg-vapor handling system. The pressure was measured with full range pressure gauges (Pfeiffer PKR\,251) at positions a, b, and~c.}
	\label{fig:1}

\end{figure}

\section*{Technical description of the mercury magnetometer}
Figure\,\ref{fig:1} shows a sketch of the mercury magnetometer test bench. 
The central part of the magnetometer is a glass cell that can be filled with \magHg{} atomic vapor. In contrast to other experiments\,\cite{Chibane1990,Romalis2004} the cell was not sealed off, but remained connected via two valves to the vacuum system and a mercury vapor source.
It was placed in the center of a three-layer cylindrical magnetic shield without end caps, which provided a shielding factor of $\sim$~500 along the axis and $\sim$~2000 perpendicular to it.
One circular polarized light beam along the cylinder axis was used for both spin-pumping and detection of the spin-precession.
For optical spin-pumping a magnetic field $B_0$, aligned with the light direction, was generated with a solenoid wound on a polyvinyl chloride~(PVC) cylinder inside the innermost shield.
The coil has a length of \unit[1600]{mm} and consists of 760 windings.
A \unit[1.6]{mA} current was used to generate a field of \unit[1]{\muT} with a measured inhomogeneity of $\Delta B/B_0\approx 3\!\times\!10^{-3}$.
A magnetic field of \unit[50]{nT} was measured with the coil turned off.
After optical pumping the spins were aligned with the pumping field $B_0$ and had to be flipped to measure a Larmor precession.
To do this we turned the main magnetic field direction non-adiabatically by almost $\pi/2$ by rapidly ramping up a second, static field $B_1\approx \unit[4]{\muT}$, perpendicular to $B_0$. The coil producing this field is a saddle coil consisting of ten frames, with three windings each, mounted on the same PVC tube as the $B_0$-coil.
After the $B_1$ field was turned on, the spins started precessing in the effective field $\vec{B}_\text{eff} = \vec{B}_0 + \vec{B}_1$ that had a
measured inhomogeneity of $\Delta B/B_\text{eff}\approx 2\!\times\!10^{-3}$ within the cell region.

\subsection*{Optical pumping}
We used the same \lampHg{} discharge UV ($\lambda \approx \unit[254]{nm}$) lamp for optical pumping and readout. A quartz bulb filled with \lampHg{}
vapor was mounted inside a microwave-cavity driven by a microwave generator such that a continuous
plasma discharge was maintained. The cavity was mounted on a three
axis micrometer table for fine adjustment of the spatial position.
The light source was mounted at the focus of a lens with a focal length of \unit[15]{cm} to obtain a nearly parallel light beam.
Both, optically spin pumping the \magHg{} vapor and reading the precession signal needs circularly polarized
light.
The light was linearly polarized with a Brewster thin-film polarizer\footnote{Brewster plate from Eksmaoptics: $\Theta_\text{b}=\unit[55]{\degree}$, extinction factor $\eta = 1/100$ (see http://www.eksmaoptics.com).}
by reflection under the Brewster angle $\Theta_\text{b}$. A quarter-wave
plate in the light path between the Brewster plate and the mercury vapor cell turned the linearly polarized light into circularly polarized. The quarter-wave
plate could be adjusted to $\pm \unit[1]{\degree}$ to the polarization plane of the linear polarized light giving a circular polarization of the light larger than \unit[96]{\%}.

All optical elements have a diameter of \unit[50]{mm} and could be adjusted relative to the position of the \magHg{}
cell inside the magnetic shield.
The light was detected by a photomultiplier\footnote{Hamamatsu R431S}~(PM), operated
at \unit[400]{V} at the far end of the shield. 

Before operation, the whole system was evacuated to pressures of $\unit[1\!\times\!10^{-5}]{mbar}\!<\!p_\text{b,c}\!<\!\unit[2\!\times\!10^{-5}]{mbar}$ and
$p_\text{a}\!<\!\unit[1\!\times\!10^{-6}]{mbar}$ over two days (see Fig.\,\ref{fig:1} for gauge positions).
During operation mercury vapor was delivered from the \magHg{} source via the inlet tube to the glass cell.
The inlet valve was typically opened for two seconds to fill \magHg{} vapor into the cell.
In standard operation the outlet valve opened after the observation of the precession signal
and mercury was pumped out of the cell. One such cycle lasted between two and three minutes.
Approximately \unit[10]{\%} of the \magHg{} vapor remained in the cell.
For safety considerations, the mercury which was pumped from the cell was trapped in a mercury filter system made of activated charcoal with sulfur additives.

\begin{figure}
\centering
	\includegraphics[width=0.9\linewidth]{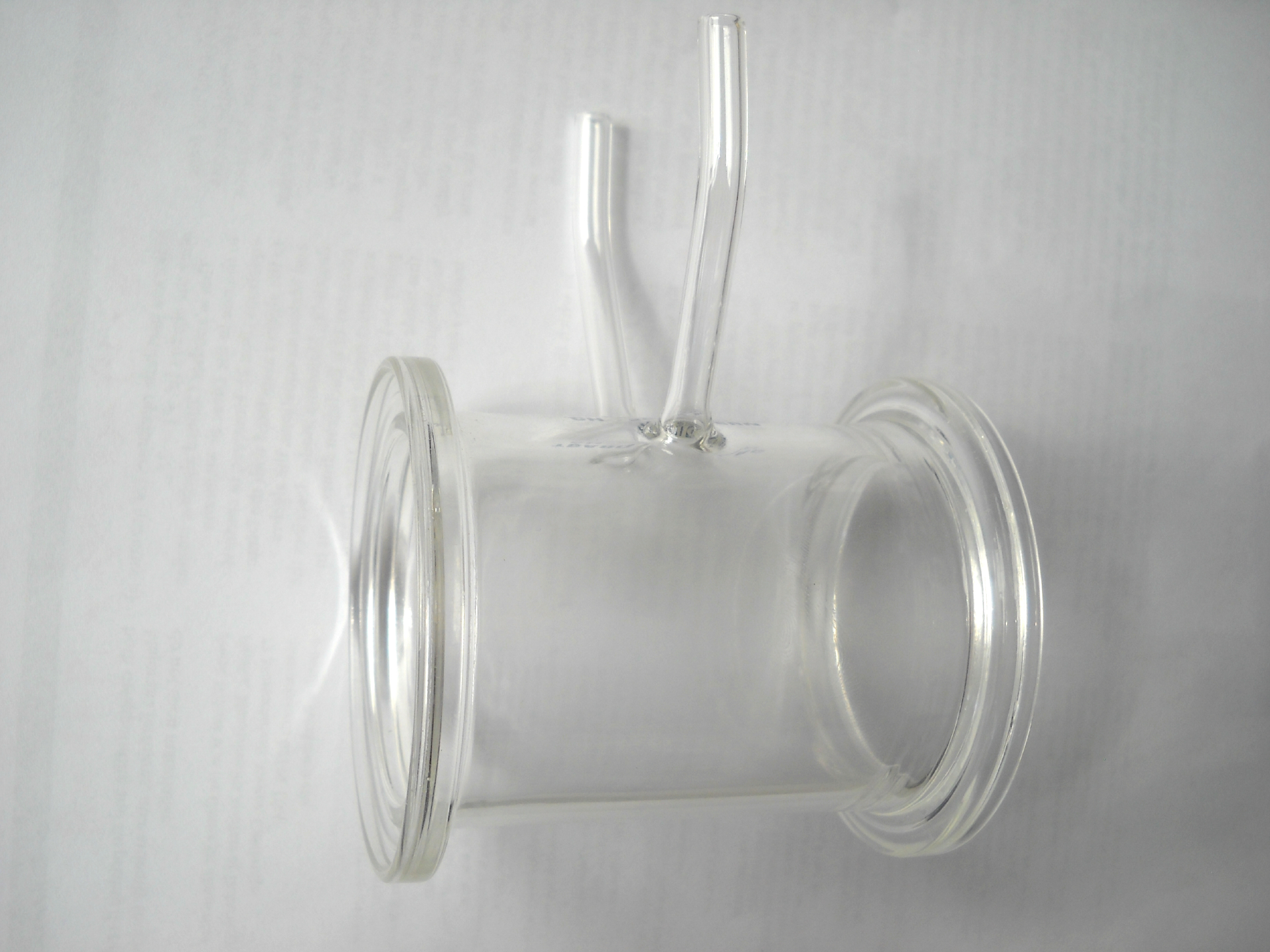}
	
	\caption{Glass cell used in the setup. O-rings in machined grooves compressed by the quartz plates guarantee vacuum tightness at both ends (not shown on the image). The two glass tubes connecting the cell to the inlet and outlet valve and are coated with Fomblin oil in all measurements.}
	\label{fig:2}

\end{figure}

The cylindrical magnetometer cells (ID = \unit[50]{mm}; $ l = \unit[80]{mm}$) are made of borosilicate glass (see Fig.\,\ref{fig:2})
and are aligned with the shield axis when installed. The end windows are circular quartz plates (OD = \unit[60]{mm},
$d = \unit[3]{mm}$) with a UV transmission of \unit[90]{\%}.
A plastic frame compresses O-rings for vacuum tightness between cylinder and end plates. The O-rings (string diameter \unit[1.7]{mm}) were placed in a \unit[1]{mm} deep, \unit[2]{mm} wide groove on the rim of the cylinder. 
In the center of the cell two glass tubes (ID = \unit[4]{mm}; $l = \unit[50]{mm}$) connect the cell to non-magnetic polytetrafluoroethylene~(PTFE) valves. The exact position along the $z$-axis of the cell is
given by the length of the aluminum tubes that connect the vacuum/Hg vapor system to the cell.

\subsection*{Data acquisition}
\begin{figure}
\centering
	\includegraphics[width=0.9\linewidth]{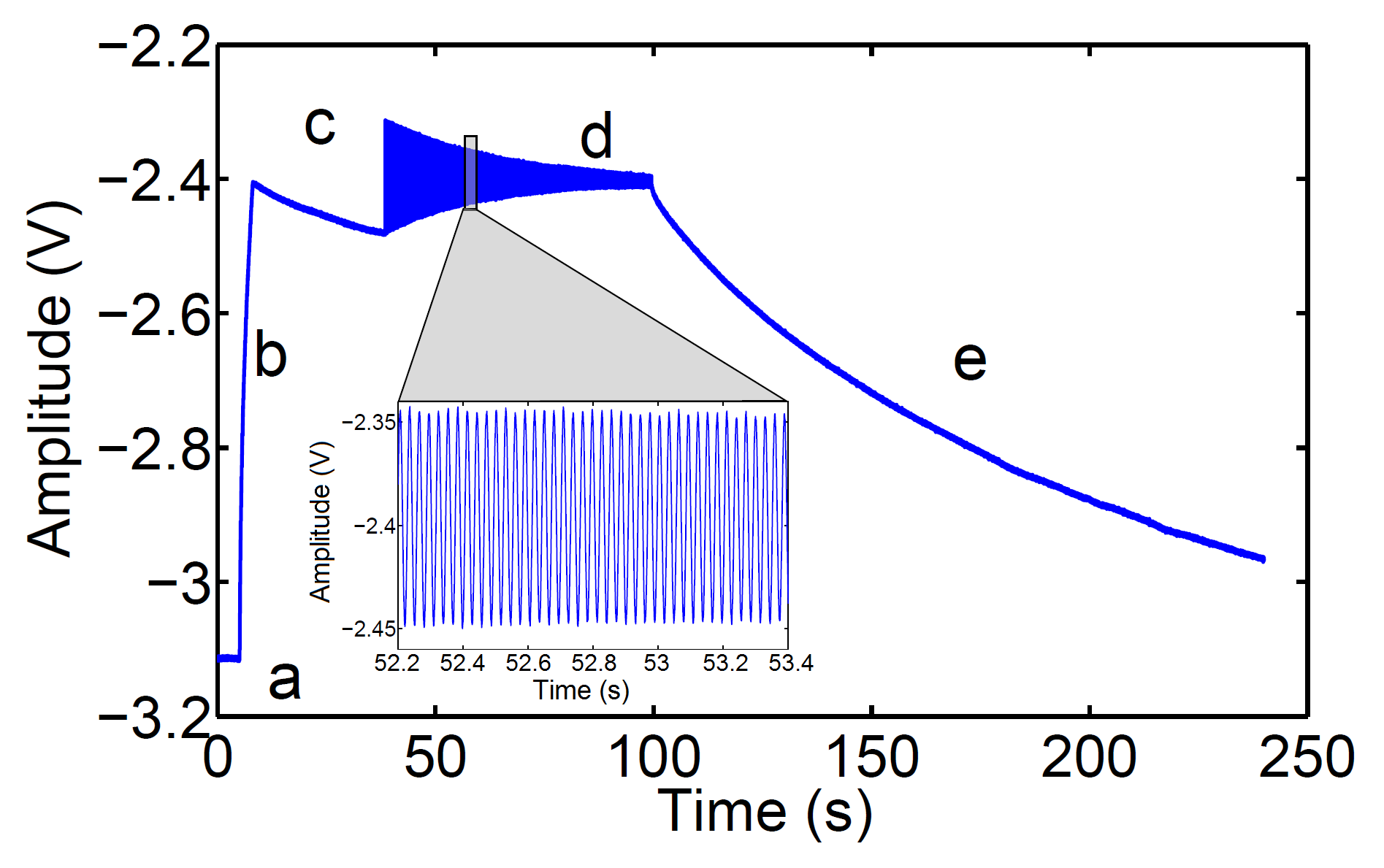}
	
	\caption{Variation of the photomulitplier signal versus time during one cycle, reflecting the five steps: (a)~measurement of the noise level for \unit[5]{s}, (b)~chamber filled with mercury (rapid change of DC-level from $\unit[-3.1]{V}$ to $\unit[-2.4]{V}$), (c)~polarization of the mercury (slow decrease of the DC-level over \unit[30]{s}), (d)~free precession of \magHg{} over \unit[60]{s}, (e)~removal of mercury from the cell. The inset shows the oscillating structure of the signal.}
	\label{fig:3}

\end{figure}

A standard National Instruments NI-USB6211 DAQ card
with a 16-bit ADC and a sampling rate of \unit[250]{kS/s} was used for data acquisition. The signal current of the PMT was grounded over a resistance
$R = \unit[1]{M \Omega}$. The voltage over this resistance was sampled at \unit[1]{kHz} and recorded. The card's digital output channels were used for controlling the vacuum valves.
One analog output channel was used to supply the current for the $B_0$ coil. The $B_1$ coil was powered by a separate current supply\footnote{PL 303-P from Thurlby Thandar Instruments Ltd.}. The card and the current supply were controlled by a LabView application. The application was structured in cycles consisting of five steps, which are explained in Fig.\,\ref{fig:3}. The duration of each step could be changed as well as the values for the currents in the coils; normally the currents remained unchanged. The amount of mercury in the cell was controlled by the opening time of the inlet valve.

\section*{Anti-relaxation coatings}
Very often the most important depolarization mechanism for \magHg{} atoms are wall collisions. Most likely, dipolar interactions at paramagnetic sites on the surface lead to depolarization of \magHg{} vapor\,\cite{Romalis2004}.
In order to keep the depolarization rate small, the surface is coated with a non-polar material. So far, Fomblin oil of type~``Y'' has
been used as anti-relaxation coating for the polarization cell of the nEDM experiment at PSI\@. Fomblin is a perfluorinated polyether consisting of carbon, fluorine, and oxygen atoms.
The C--F bonds result in low vapor pressures and the oil is normally used in vacuum pumps.
%
%

In the past positive results were reported
for non-polar paraffin or paraffin-like coatings\,\cite{Romalis2004} which can be considered as standard surface coatings in many experiments dealing with polarization of gases and vapors. From our own experience we knew that perfluorinated hydrocarbons like PTFE or Fomblin oil are good anti-relaxation coatings, hence we also
studied a perfluorinated paraffin from Maflon\footnote{Maflon: http://www.maflon.com (2010)}.
The materials we tested in our setup are listed in Tab.\,\ref{Tab:1}. AquaSil\footnote{AquaSil$^{\text{\textregistered}}$ is a registered trademark of Dentsply.} and SurfaSil\footnote{SurfaSil$^{\text{\textregistered}}$ is a registered trademark of Siltech LLC.} are hydrocarbons containing silicon atoms that react directly with glass and have been reported as potential candidates in Ref.\,\cite{Chibane1990}. 
These solid coatings could make re-coating of the cell unnecessary. 
The ``J'' oil from Apiezon\footnote{Apiezon$^{\text{\textregistered}}$ is a registered trademark of M\&I Materials.} was chosen because it
is a well-known vacuum oil that is based on hydrocarbons.
Finally, we tested black PTFE, a mixture
of PTFE with \unit[25]{\%} carbon to color the material black. We expected that the excellent anti-depolarization
properties of PTFE would be maintained while the black color would significantly reduce the reflection of
linear polarized light re-emitted from the \magHg-atoms during the polarization process. However, it turned out to be completely unsuitable.

\subsection*{Coating the glass cell}

In all cases the glass cells were first cleaned with water and soap, rinsed with demineralized water and then dried with a hot air gun. 
Fomblin oil and grease, as well as the ``J'' oil of Apiezon were uniformly spread on the inside of the cell with a nitrile-glove-covered finger.

AquaSil and SurfaSil were applied by immersion of the glass cells into a solution of these chemicals.
AquaSil was purchased as a \unit[20]{\%} solution in alcohol which was further diluted in water to a \unit[1]{\%} solution. SurfaSil was diluted in acetone to a \unit[1--10]{\%} solution.
In both cases the cells were immersed for five to ten seconds, during which the solutions were continuously stirred with a magnetic stirrer. Afterwards, the cells were rinsed thoroughly with methanol to remove the excess, unreacted solution and were dried for \unit[24]{h} in a fume hood.

\begin{figure}%
\centering
	\includegraphics[width=0.7\columnwidth]{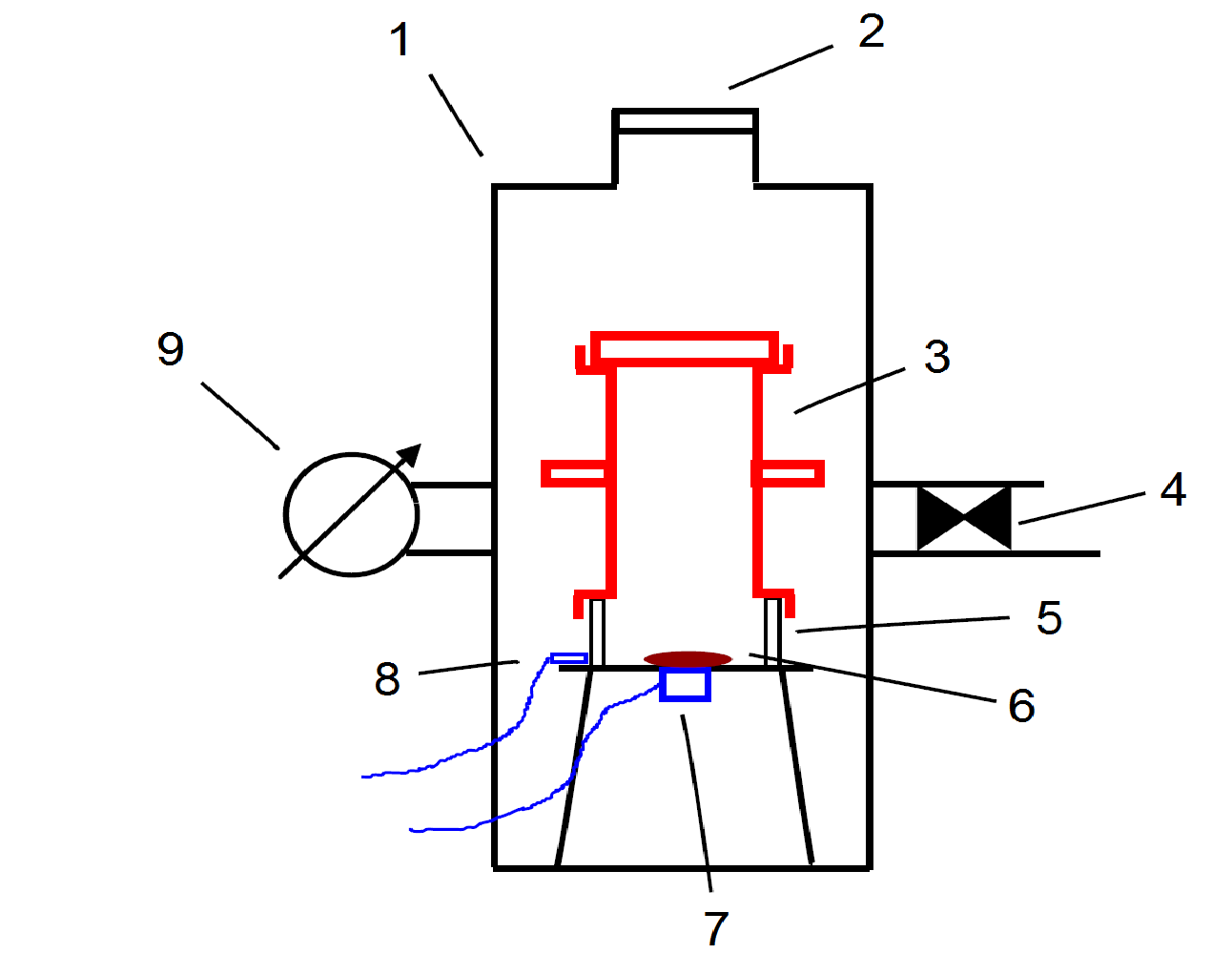}%
\caption{Setup for vacuum deposition of normal and perflourinated paraffin. 1)~Vacuum vessel, 2)~window flange to observe progress of evaporation, 3)~glass cell with one quartz plate, 4)~vacuum valve to pumping unit 5)~steel tube for thermal insulation, 6)~paraffin, 7)~resistive heater, 8)~thermometer, 9)~pressure gauge.}%
\label{fig:EvaporationSetup}%
\end{figure}

The normal and perflourinated paraffin coatings were applied by evaporation (fig.\,\ref{fig:EvaporationSetup} shows the evaporation setup).
%
%
The cells were placed upright on a copper heating plate inside a vacuum vessel. 
Instead of putting the cell directly on the heating plate, we used a stainless steel ring to provide a thermal insulation and thus help keep the glass cool. In order to obtain an even coating we repeated the coating process with the cell turned up side down.
A small amount ($<\unit[0.1]{g}$) of paraffin was placed inside the stainless steel ring on top of the copper plate. The cell was then closed off with one of the quartz windows on top.
Once the vessel was under vacuum, the heater was turned on to melt the paraffin and start the coating process. Ordinary paraffin melts at $\sim \unit[80]{\celsius}$ while perflourinated paraffin sublimes at $\sim \unit[160]{\celsius}$.
%
A window on the vacuum vessel allowed us to observe and control the deposition.
After $\sim \unit[20]{min}$ the paraffin flake had disappeared and the cell was coated. 

In all cases the quartz end windows were coated with the same material during the same process; the two connection tubes were always coated with Fomblin oil and the O-rings with Fomblin grease.

\subsection*{Measurement of the decay time constants}

The depolarization rate of a precessing population of spin polarized atoms is a combination of
depolarization due to wall collisions and inhomogeneous magnetic fields:

\begin{equation}
	1/\tau = 1/T_2 + 1/\tau_\text{w}.
\label{eq:3}
\end{equation}

\noindent Here $T_2$ is the transversal polarization decay time constant due to magnetic field inhomogeneities and $\tau_\text{w}$ is the decay time constant due to wall interactions. Measured magnetic field maps helped us to estimate $T_2\approx \unit[200]{s}$ using equation~(62) in Ref.\,\cite{Cates1988PRA37}. The experiment was performed at $\sim \unit[22]{\celsius}$ which results in a mean velocity $v_\text{Hg} \approx \unit[177]{m/s}$ and an average wall collision frequency of $r\approx \unit[4.7]{kHz}$ in the cells.

We measured the relaxation time with an adapted version of the ``relaxation in the dark'' method
pioneered by Franzen\,\cite{Franzen1959}\footnote{Franzen originally measured the longitudinal depolarization rate $T_1$ by measuring the increase of the transmitted light intensity after blocking the pump light beam.}. In this method \magHg{} was filled into the chamber and the atoms were optically pumped to the maximum polarization, which was indicated by a maximum light transmission through the magnetometer cell.
Then we non-adiabatically switched on $B_1$ and blocked the reading light for a ``dark time'' of $\Delta t = \unit[5\dots60]{s}$, after which the amplitude of the precession signal was measured. Typically, the initial amplitude after \unit[5]{s} ``dark time'' had a signal/noise ratio of $\sim 50$.
Finally the cell was pumped for about \unit[100]{s} to remove the mercury.
It is important that the whole measurement was performed with the same \magHg{} density, since the polarization depends slightly on the number density.
Ideally one would have filled the cell only once with mercury and never opened the valves during the measurement. However, then the residual pressure would have risen due to out-gassing and small leaks and collisions with residual gas molecules would have distorted the measurement.

\begin{figure}
\centering
	\includegraphics[width=0.9\linewidth]{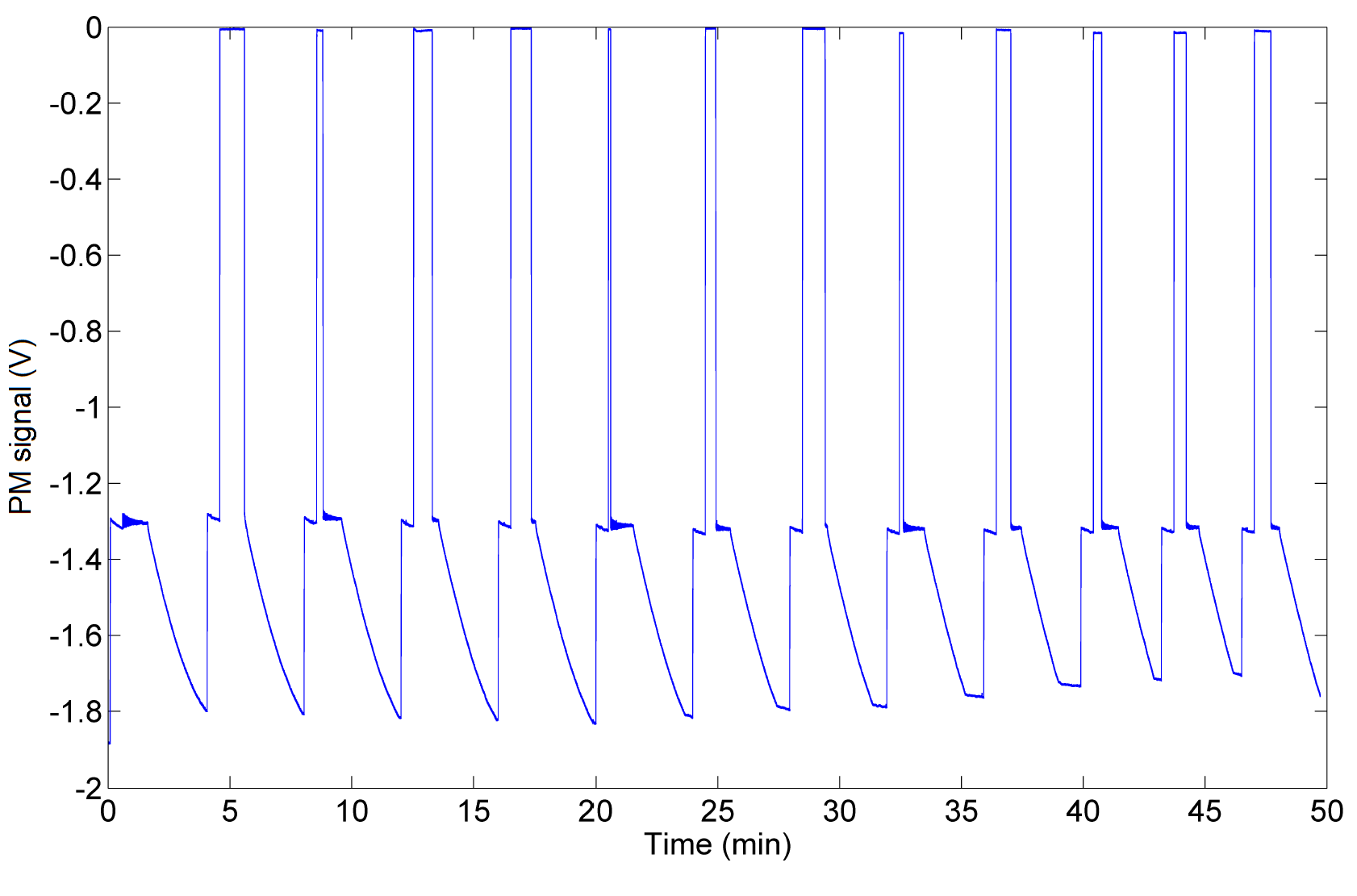}
	
	\caption{The plot shows  the photo multiplier signal in volts during several polarization and free precession cycles with varying ``dark times''. In each cycle one can distinguish the initial filling of the cell with Hg vapor (sharp signal change from \unit[-1.8]{V} to \unit[-1.3]{V}); the polarization of the vapor; the blocking of the light for variable ``dark times'' between  \unit[0]{s} and \unit[60]{s} (PM signal at \unit[0]{V}), the free precession after removing the light block and the final vacuum pumping of the cell (slow signal decrease from \unit[-1.3]{V} to \unit[-1.8]{V}). From the ``dark time'' dependent change of the initial free precession amplitude after blocking the readout light we deduced the depolarization time constant $\tau$.}
	\label{fig:4}

\end{figure}

This method combines two advantages: ({\it i}) we do not have to worry about the influence of the read out light, which might have significantly depolarized the \magHg{} ensemble, and ({\it ii}) measuring the amplitude and not the absolute light intensity after the ``dark time'' makes the method more robust against intensity drifts of the lamp.
An example of a measurement is shown in Fig.\,\ref{fig:4}.

\begin{figure}
\centering
	\includegraphics[width=0.9\linewidth]{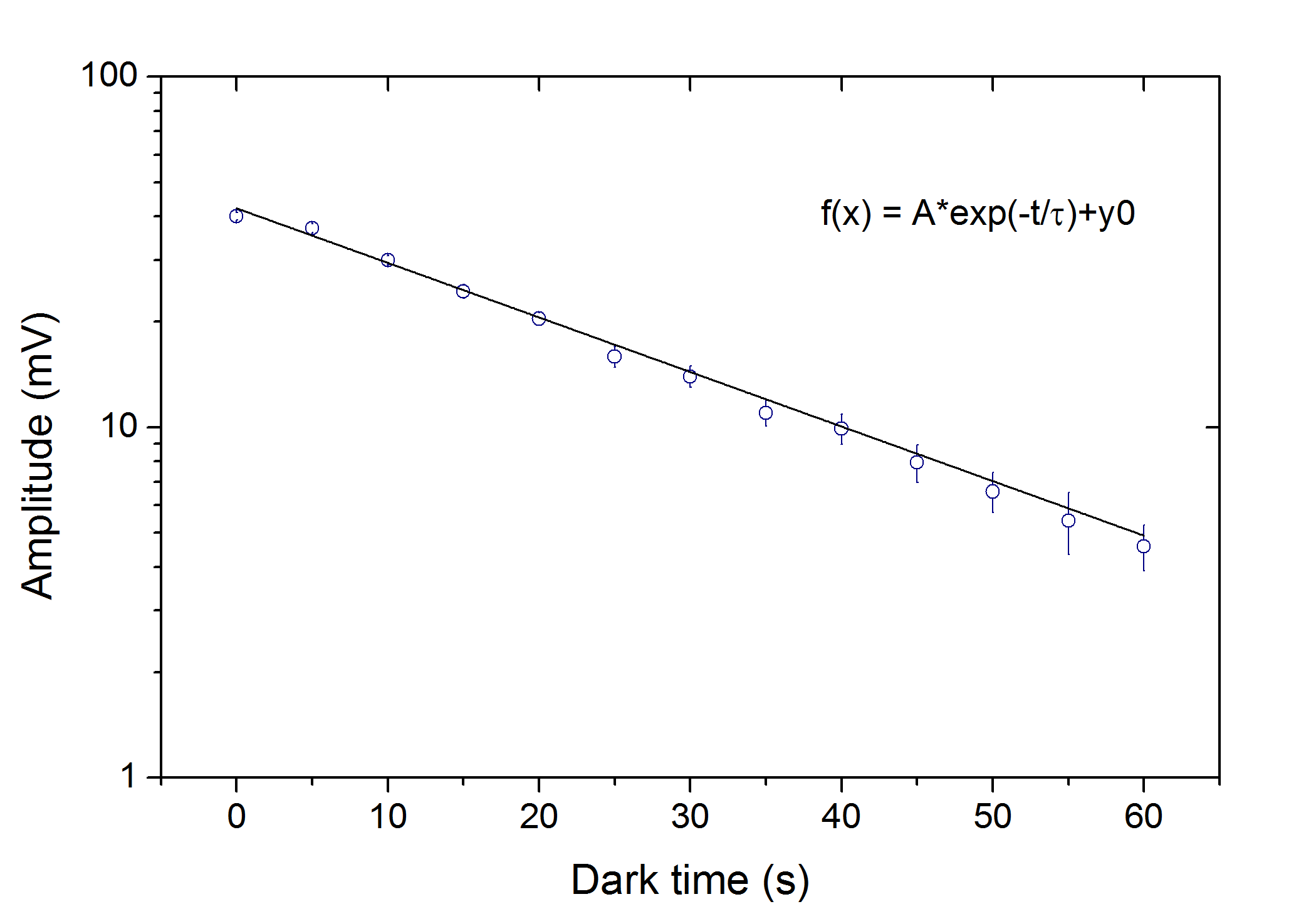}
	
	\caption{The amplitude of the signal as a function of the dark time $t_\text{dark}$. The decay constant $\tau$ was extracted from an exponential fit to the data. The example graph shows a measurement of paraffin coating: $\tau = \unit[28(2)]{s}$, $A =\unit[42(1)]{mV}$, $y0=0(2)$ with a reduced $\chi^2 = 1.22$.}
	\label{fig:5}

\end{figure}

The relaxation time, $\tau$, was determined by fitting a single exponential (see Fig.\,\ref{fig:5}) to initial precession amplitudes extracted for different times in the dark. 
Each coating was measured twice with a time difference of at least two weeks between the measurements. During this time the cells remained untouched and exposed to normal air conditions. The results are summarized in Tab.\,\ref{Tab:1}. 
From the measured values it is clear that most of the tested materials are as good as, or even slightly better than, the previously used Fomblin oil. The slight decrease in $\tau$ in some of the second measurements might be interpreted as a degradation of the surface quality with time or as a change of the magnetic field conditions. 
The longest $\tau$ was measured with perfluorinated paraffin from Maflon, a skiing wax. The ``J'' oil of Apiezon was found to be highly magnetic, as measured with a fluxgate magnetometer, which explains why no signal was seen. Black PTFE showed a very short $\tau$ less than \unit[5]{s} and was not measured a second time. A possible reason might be contaminants in 
the material which cause the \magHg{} atoms to depolarize.
However, we found no signal ($\Delta B\!<\!\unit[20]{pT}$ at a distance of \unit[3]{cm}) indicative of a magnetic contamination in a measurement with a cesium-magnetometer-based gradiometer\,\cite{Groeger2005,Pazgalev2008}.

\begin{table}
	\centering
		\begin{tabular}{l|c|c}
		\hline
		Coating material and its chemical structure & $\tau_1$\,(s) & $\tau_2$\,(s) \\
		\hline\hline
		
		Perfluorinated paraffin & \multirow{2}{*}{$40.5(1.2)$} & \multirow{2}{*}{$34.6(6)$} \\
		\small{\ce{C_{20}F_{42}}(\unit[80]{\%}) + \ce{C_{16}F_{34}}(\unit[10]{\%}-\unit[20]{\%})}  & & \\ \hline
		
		Fomblin grease & \multirow{3}{*}{$34.7(7)$} & \multirow{3}{*}{$30.6(5)$} \\
		\small{\ce{CF_3-[-O-CF_2-CF_2]_n-[O-CF_2]_m-O-CF_3}} & & \\
		\footnotesize{mixed with \ce{CF_2-CF_2}} & & \\ \hline
		
		SurfaSil & \multirow{2}{*}{$32.2(4)$} & \multirow{2}{*}{$30.6(7)$} \\
		\small{\ce{Cl-[-(CH_3)_2-Si-O-]_n-(CH_3)_2-Si-Cl}}  & & \\ \hline
		
		Paraffin & \multirow{2}{*}{$26.3(4)$} & \multirow{2}{*}{$28.0(2)$} \\
		\small{\ce{C_{32}H_{66}}}  & & \\ \hline	
		
		Fomblin oil type ``Y''& \multirow{3}{*}{$28.4(5)$} & \multirow{3}{*}{$27.0(6)$}\\
		\small{\ce{CF_3-[-O-CF_2-CF_2]_n-[O-CF_2]_m-O-CF_3}} & & \\
		\footnotesize{n/m = $20 \dots 40$} & & \\ \hline
				
		AquaSil & \multirow{2}{*}{$23.0(1.0)$} & \multirow{2}{*}{$26.0(6)$} \\
		\small{\ce{CH_{3}-(CH_2)_{15}-Si-(OH)_{3}}}  & & \\ \hline
		
		Black Teflon & \multirow{2}{*}{$<5$} & \multirow{2}{*}{--} \\
		\small{\ce{CF_2-CF_2} mixed with carbon}  & & \\ \hline
		
		Apiezon ``J'' oil & \multirow{2}{*}{no signal} & \multirow{2}{*}{--} \\
		\footnotesize{based on hydrocarbons} & & \\ \hline	
		
		\end{tabular}
	\caption{Measured $\tau$-times (first measurement $\tau_1$, second $\tau_2$) for studied coatings. The error of the time constants are fit errors scaled for reduced $\chi^2 =1$. The $\tau$-times measured inside the neutron precession chamber of the nEDM apparatus at PSI are typically $\sim \unit[100 - 180]{s}$. It is made of two diamond-like-carbon~(DLC) coated electrodes and a deuterated polysterol~(dPS)\,\cite{Bodek2008} coated insulator ring. Note that the neutron precession chamber is much larger and has a mean free path of \unit[19]{cm}, and hence an average wall collision frequency of \unit[930]{Hz}, giving similar depolarization rates as in these measurements.}
	\label{Tab:1}
\end{table}

\section*{Conclusion and discussion}

A \magHg{} magnetometer test apparatus was built to investigate potential improvements of the \magHg{} magnetometer of the nEDM experiment at the Paul Scherrer Institute.
As a first study, depolarization times of different magnetometer cell coatings were measured.
Several materials showed similar or even longer relative depolarization times than the previously used Fomblin. Previous measurements of Fomblin had an average depolarization probability per wall collision of $\beta_\text{Fomblin} =3.5(5)\!\times\!10^{-6}$\,\cite{RocciaPhD} which is within a factor of two of our measurement. A similar discrepancy has been noted for the Fomblin grease measurement where previous studies gave depolarization values of $\beta_\text{FomGr}=2.1\!\times\!10^{-6}$\,\cite{MayPhD}, also approximately a factor of two smaller. 
These discrepancies might be explained with depolarization upon collisions with the fluoroelastomer o-rings which were used to seal off the cell, or via collisions with the inner surfaces of the two PTFE valves. For all materials we took great care that the measurements were performed under identical conditions. Small  variations in the magnetic field environment, which was not monitored during measurements, might have slightly changed the $T_2$-time but cannot explain a factor of two. The depolarization time of an uncoated cell was not measured as our experience with the mercury magnetometer inside the nEDM experiment showed worse performances for partially uncoated cells. 

Perfluorinated paraffin and SurfaSil were found to be the most interesting candidates and will be further investigated.
Both have the important advantage that the coating is of long durability and re-coating during nEDM measurements might become unnecessary in the future. 
In general the perfluorinated materials seem to allow for longer depolarization times than hydrogenated materials. This might be explained by a more pronounced dipolar character of hydrogen bonds. The long term stability of the solid coatings remains to be investigated. In future studies using the test apparatus the magnetic field monitoring and control will be improved.

\begin{acknowledgement}
The authors would like to thank the University of Fribourg atomic physics group for the preparation of \lampHg{} bulbs and help with the cesium magnetometry. We thank the initial support of this work by TU M\"{u}nchen, in particular by P.~Fierlinger and G.~Petzoldt. The authors are grateful for the excellent technical support by F.~Burri and M.~Meier.
\end{acknowledgement}


\end{document}